\documentclass[aps,prl,twocolumn,groupedaddress,floatfix]{revtex4}

\usepackage{graphicx}
\usepackage{dcolumn} 
\usepackage{bm}      

\begin{document}

\title{p-wave Feshbach molecules}
\author{J. P. Gaebler}
\email[Electronic address: ]{gaeblerj@jila.colorado.edu}
\homepage[URL: ]{http://jilawww.colorado.edu/~jin/}
\author{J. T. Stewart}
\author{J. L. Bohn}
\author{D. S. Jin}

\affiliation{JILA, Quantum Physics Division, National Institute of
Standards and Technology and Department of Physics, University of
Colorado, Boulder, CO 80309-0440, USA}

\date{\today} \begin{abstract}
We have produced and detected molecules using a p-wave Feshbach
resonance between $^{40}$K atoms. We have measured the binding
energy and lifetime for these molecules and we find that the binding
energy scales approximately linearly with magnetic field near the
resonance. The lifetime of bound p-wave molecules is measured to be
$1.0 \pm 0.1$ ms and $2.3 \pm 0.2$ ms for the $m_l = \pm 1$ and $m_l
= 0$ angular momentum projections, respectively. At magnetic fields
above the resonance, we detect quasi-bound molecules whose lifetime
is set by the tunneling rate through the centrifugal barrier.
\end{abstract}

 \pacs{??)}

\maketitle


Much recent work with atomic Fermi gases has taken advantage of the
ability to create strong atom-atom interactions through the use of
magnetic-field Feshbach resonances. A Feshbach resonance occurs when
the energy difference between a diatomic molecule state and two
scattering atoms can be tuned to zero. This energy difference can be
tuned with a magnetic field if there is a difference between the
magnetic moment of the molecule and that of two free atoms.
Experiments have taken advantage of these tunable interactions to
study atom pair condensates in the BCS superfluid to Bose-Einstein
condensate crossover regime\cite{Regal2003a, Jochim2003a,
Regal2004a, Zwierlein2005a}. While this work involved pairing with
s-wave interactions, there are several reasons why the study of
possible atom condensates with non-s-wave pairing is compelling. For
a p-wave paired state, the richness of the superfluid order
parameter leads to a complex phase diagram with a variety of phase
transitions as a function of temperature and interaction strength
\cite{Gurarie2005a,ChiHo2005,Ohashi2005,Botelho2005}. Some of these
phase transitions are of topological nature \cite{Read2000}, and
have been predicted to be accessible via detuning between the BCS
and BEC limits with a p-wave Feshbach resonance. Furthermore,
because p-wave resonances are intrinsically narrow at low energies,
due to the centrifugal barrier, a quantitatively accurate
theoretical treatment is possible \cite{Gurarie2005a}.

p-wave resonances have been observed in Fermi gases of $^{40}$K
atoms \cite{Regal2003c,Gunter2005a} and $^{6}$Li atoms
\cite{Zhang2004a,Schunck2005} via measurements of elastic scattering
and inelastic loss rates. There is also suggestive evidence for
molecule creation in $^{6}$Li using magnetic-field sweeps through
the resonance \cite{Zhang2004a}. However, it is still not known
whether long-lived Feshbach molecules with nonzero angular momentum
can be created from a gas of fermionic atoms.

In this Letter, we present evidence for the production and direct
detection of p-wave molecules created with a p-wave Feshbach
resonance. We measure the lifetimes and binding energies of these
molecules and perform theoretical calculations to interpret our
results.  We are able to extend our measurements to the p-wave
quasi-bound state.

\begin{figure}
\includegraphics[width=\linewidth]{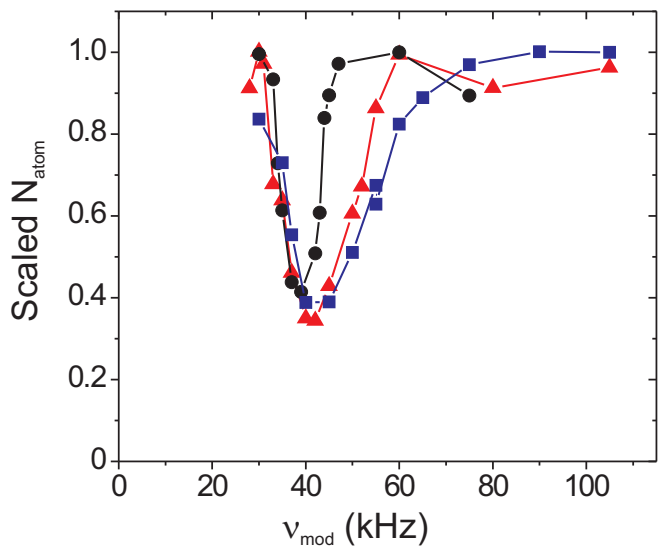}
\caption{\label{lineshapes} (color online) Line shapes for
association to a bound p-wave molecule using a sinusoidally
modulated magnetic field.  For this data, $B = 198.12$ G.  We find
that the line shape is asymmetric and has a width that depends on
the cloud energy $E_G$, as defined in the text.  The data are for
values of $E_G$ of $2.0$ kHz (circles), $4.5$ kHz (triangles), and
$6.0$ kHz (squares). To highlight the asymmetry and dependence on
$E_G$, the total number of atoms, $N_{atom}$, for each line shape
was scaled to one.}
\end{figure}

\begin{figure}
\includegraphics[width=\linewidth]{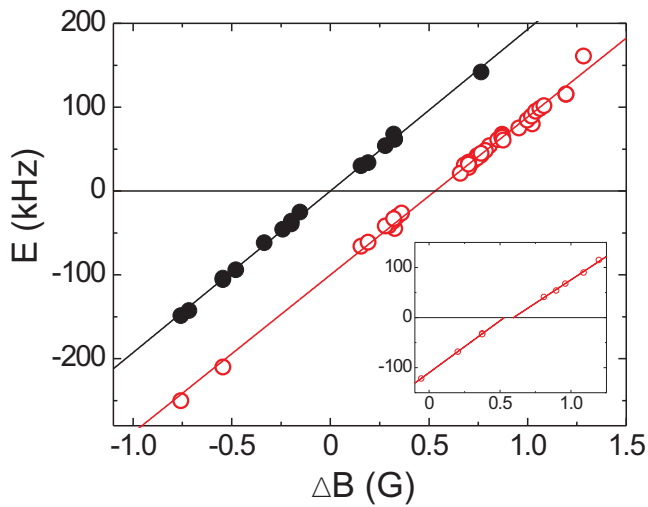}
\caption{\label{bindingenergy} Energy of the molecule as a function
of magnetic field for both the $m_l=0$ (open circles) and $m_l=\pm1$
(closed circles) resonances.  $E<0$ corresponds to a bound molecule;
$E>0$ corresponds to a quasi-bound state. $\Delta B$ is the detuning
from the $m_l = \pm1$ resonance position, which we measure to be $B
= 198.3\pm.02$ G.  Linear fits give a slope of $188\pm2$ kHz/G for
the $m_l=0$ resonance and $193\pm2$ kHz/G for $m_l=\pm1$. The inset
shows data for the $m_l=0$ resonance that suggest some non-linearity
near the resonance. These data were taken all in one day to reduce
the uncertainty in the magnetic field.}
\end{figure}

Our experiments were carried out using a quantum degenerate $^{40}$K
gas near a p-wave resonance between atoms in the $|f,m_f \rangle =
|9/2,-7/2 \rangle$ spin state, where $f$ is the total atomic spin
and $m_f$ is the magnetic quantum number. We cool a mixture of
$|9/2,-7/2 \rangle$ and $|9/2,-9/2 \rangle$ atoms to quantum
degeneracy in a crossed-beam optical dipole trap using procedures
outlined in previous work \cite{Regal2005c}. The trap consists of a
horizontal laser beam along $\hat{z}$ with a $\frac{1}{e^2}$ radius
of 32 $\mu$m and a vertical beam along $\hat{y}$ with a
$\frac{1}{e^2}$ radius of 200 $\mu$m.  A magnetic-field points along
the $\hat{z}$ direction and is held at $B = 203.5$ G during the
final stage of evaporation. The final conditions consist of $10^5$
atoms per spin state at a temperature of $T\approx 0.2
\hspace{0.05cm}T_F$, where $T_F = E_F/k_b$ is the Fermi temperature
and $k_b$ is Boltzmann's constant. The final trap frequencies are
typically $\omega_r/2\pi=180$ Hz and $\omega_z/2\pi=18$ Hz, yielding
$E_F/h \approx 7$ kHz. To probe the atom cloud, we turn off the
trapping potential, wait for a variable expansion time of 1.9 to 10
ms, and then send a $40$ $\mu$s pulse of resonant laser light
through the cloud along $\hat{z}$ onto a CCD camera.  We expect that
this resonant absorption imaging should only be sensitive to atoms,
and not to p-wave molecules.

To measure the binding energies of the p-wave molecules, we
resonantly associate the atoms into molecules using a sinusoidally
modulated magnetic field. This method has been previously used to
both dissociate and associate s-wave Feshbach molecules
\cite{Greiner2005a,Thompson2005a}.  We ramp the magnetic field to a
value near the Feshbach resonance and then apply a small sinusoidal
oscillation at a frequency $\nu_{mod}$ for a duration of 36 ms. The
amplitude of the modulation is a Haversine envelope to reduce the
power in frequencies other than $\nu_{mod}$.  As we vary
$\nu_{mod}$, we observe a resonant decrease in the number of atoms,
$N_{atom}$, in the $|9/2,-7/2 \rangle$ state.  Sample data sets are
shown in Fig. \ref{lineshapes}.  We interpret the loss of atoms to
be caused by resonant association to the molecular state.

The line shapes we observe for association to bound states are
asymmetric and have widths that increase linearly with cloud energy,
$E_G$, which we obtain from the width of a Gaussian fit to an
expanded cloud. These line shape features are characteristic of a
narrow resonance where the energy width of the atom cloud dominates
over any intrinsic width of the resonance.  The asymmetry of the
line shape reflects the distribution of collision energies in the
Fermi gas. For our measurements we used a range of modulation
amplitudes of $100$ to $500$ mG and did not observe any significant
amplitude dependent broadening or shifting of the line shapes.
However, for our largest modulation amplitudes we observe harmonics
of the principal feature.

From the magneto-association line shape we can extract the pair
energy, $E$, relative to the energy of free atoms.  Because the line
shape shifts as $E_G$ is increased, we extract the resonance
position by measuring the frequency for maximum loss at different
values of $E_G$ and extrapolating to $E_G=0$.  This gives a
correction whose magnitude is approximately $3$ kHz.  In Fig.
\ref{bindingenergy} we plot the pair energy as a function of
magnetic-field detuning.

As can be seen in Fig. \ref{bindingenergy}, there is a splitting of
$0.509 \pm.03$ G in the p-wave resonances for $m_l=0$ and $m_l =
\pm1$, where $m_l$ is the pair orbital angular momentum projection
onto the magnetic-field axis. This splitting is caused by the
magnetic dipole interaction and was first observed in Ref.
\cite{Regal2003c} and explained in Ref. \cite{Ticknor2004a}. For
both resonances, we observe a linear dependence of the pair energy,
$E$, on magnetic-field detuning.

With resonant magneto-association we find that we can associate
atoms into quasi-bound states above the resonance, as well as into
bound molecular states below the resonance.  These quasi-bound
states are paired states with positive energy and lifetimes set by
the tunneling time through the p-wave centrifugal barrier. The
height of the barrier for $^{40}$K is $h \times 5.8$ MHz, where $h$
is Planck's constant. This tunneling time causes the widths of line
shapes for association to quasi-bound states to be as much as three
times greater than for bound states, for data taken with similar
initial cloud energies.

\begin{figure}
\includegraphics[width=\linewidth]{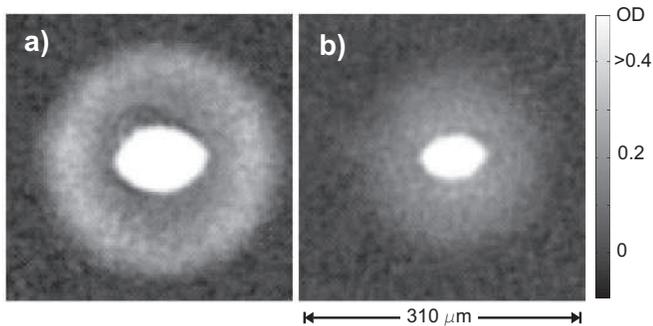}
\caption{\label{rings} (a) Dissociated $m_l=\pm1$ molecules in the
$xy$ plane.  (b) Dissociated $m_l=0$ molecules in the $xy$ plane.
The bright spot in the center corresponds to cold atoms that did not
form molecules. The color mapping for the cloud's optical density
(OD) is shown on the right. In each image the dissociation is done
with a $10$ $\mu$s linear ramp to a B-field value $850$ mG above the
respective resonance.  The expansion time before imaging is $2.5$
ms. Each image represents the average of 5-10 shots. To reduce high
frequency noise, each pixel is taken to be an average of itself and
its four nearest neighbors. The total number of atoms in the
$m_l=\pm1$ images is $50\%$ higher than for the $m_l=0$ images. The
images are consistent with the expected angular distributions of
$sin^2(\theta)$ and $cos^2(\theta)$ for the $m_l=\pm1$ and $m_l=0$
states, respectively, where $\theta$ is the angle from the $\hat{z}$
axis.}
\end{figure}

\begin{figure}
\includegraphics[width=\linewidth]{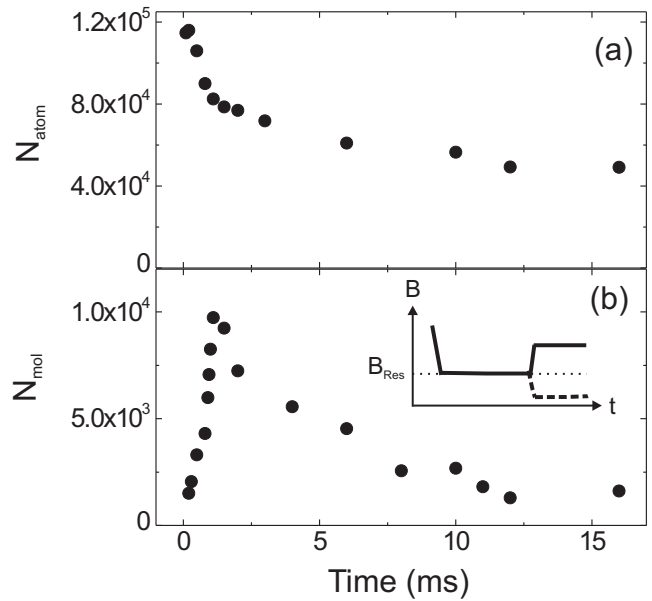}
\caption{\label{formation} (a) Measured atom number as a function of
time that the magnetic field is held at the $m_l = \pm 1$ resonance.
(b) Measured molecule number for the same hold time on resonance.
The inset shows the timing sequence for this experiment. The
magnetic field is ramped to the $m_l = \pm 1$ resonance, $B_{Res}$,
in less than $100$ $\mu$s and then held at that field for a variable
amount of time. At the end of this hold time we measure the number
of molecules in the gas using the dissociation technique described
in the text (solid line). The number of atoms not in molecules is
measured by ramping the field below the resonance where the
molecules are deeply bound and then expanding and imaging the atoms
at this field (dashed line).}
\end{figure}

We have taken advantage of the tunneling of quasi-bound pairs to
observe the molecules.  For these experiments we use a
spin-polarized gas of atoms in the $|9/2,-7/2 \rangle$ state to
eliminate non-resonant collisions.  A pure spin-polarized gas is
obtained from a $95/5$ mixture of $|9/2,-7/2 \rangle$ and $|9/2,-9/2
\rangle$ atoms by removing atoms in the spin state $|9/2,-9/2
\rangle$ with a slow sweep through an s-wave resonance at a magnetic
field $B = 202.1$ G. Near this resonance, the large inelastic loss
rates for high density clouds ensure that nearly all of the atoms in
the $|9/2,-9/2 \rangle$ spin state are lost \cite{Regal2004b}. The
magnetic field is then set to $B=199.7$ G, and the optical trap
depth is lowered to reach $10^5$ atoms. As the trap depth is lowered
to its final value, the spin-polarized Fermi gas is not able to
rethermalize because s-wave collisions are forbidden by quantum
statistics and higher-order partial wave collisions are frozen out
due to the Wigner threshold law. A Gaussian fit to an expanded cloud
yields energies of approximately $k_b \times 130$ nK, $k_b \times
70$ nK, and $k_b \times 230$ nK in the $\hat{x}$, $\hat{y}$ and
$\hat{z}$ directions, respectively.

To detect molecules in the gas, we quickly increase the magnetic
field, in 10$\mu$s, to a value above the resonance where quasi-bound
molecules have a large positive energy. The paired atoms then
quickly tunnel out of the centrifugal barrier and this energy is
converted to kinetic energy of atoms flying apart. We immediately
turn off the optical trap, expand for a variable time of $1.9$ to
$5$ ms, and take an absorption image. The result is a large
energetic cloud of atoms surrounding the cold gas, as seen in Fig.
\ref{rings}.  We note that the 10 $\mu$s magnetic-field ramp is much
shorter than any trap period or collision time scale. This is
similar to the detection scheme used in Ref. \cite{Durr2004a}. The
angular dependence in the distribution of energetic atoms is
consistent with p-wave pairing (see Fig. \ref{rings}). We have
verified that the size of the energetic cloud depends on the
magnetic-field ramp, with faster ramps to higher fields yielding
larger clouds. To place a lower limit on the molecule number,
$N_{mol}$, we can sum the atom signal occurring outside of a certain
radius surrounding the cold inner cloud.  $N_{mol}$ is $1/2$ the
number of atoms found outside this radius.

\begin{figure}
\includegraphics[width=\linewidth]{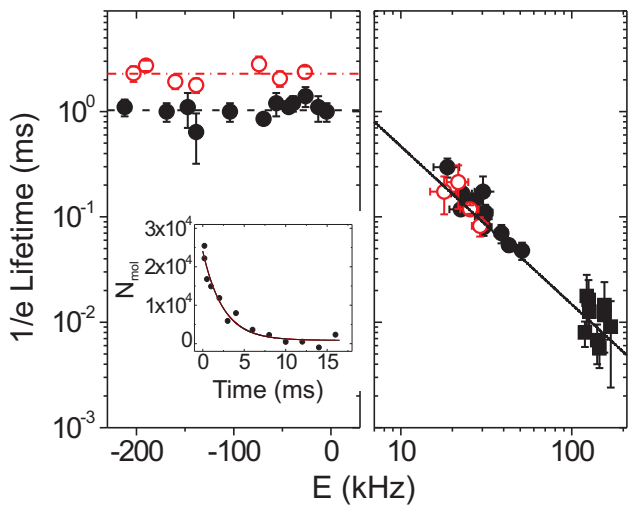}
\caption{\label{lifetimes} The lifetimes of p-wave molecules as a
function of binding energy. Circles indicate directly measured
lifetimes, while squares indicate lifetimes inferred from
magneto-association line shape widths. Data for the $m_l=0$
resonance are shown in open symbols, while $m_l=\pm1$ are closed
symbols.  The two dotted lines on the left indicate the averages of
the measured bound state lifetimes. The solid line on the right is
the theory curve for the quasi-bound lifetimes, as discussed in the
text. The inset shows a sample exponential lifetime curve of the
$m_l=0$ bound state taken at 1 G below the resonance.}
\end{figure}

We have measured the lifetimes of p-wave molecules in a
spin-polarized gas as a function of magnetic-field detuning from the
resonance.  We first create molecules by quickly ramping the
magnetic field to a value near the resonance and then holding at
that field for $1$ to $2$ ms. The span of magnetic-field values for
which we are able to observe molecule creation is approximately $90$
mG. With this technique, we can transfer nearly $20\%$ of the
initial atom population to molecules, as shown in Fig.
\ref{formation}. For a lifetime measurement, after we create
molecules, we ramp the magnetic field to a test value and hold there
for a variable amount of time before detecting the molecules. We
measure $N_{mol}$ as a function of the hold time at the test
magnetic field. A sample lifetime curve is shown in the inset of
Fig. \ref{lifetimes}.  We apply this technique to directly measure
the lifetime of both the bound and quasi-bound states. The results
of these lifetime measurements (circles) are shown in Fig.
\ref{lifetimes}. Also shown in Fig. \ref{lifetimes} are lifetimes
extracted from the widths of magneto-association line shapes on the
quasi-bound side (squares) \cite{widths}. On the quasi-bound side,
we find that the pair lifetime follows the expected $\tau \propto
E^{-\frac{3}{2}}$ behavior for tunneling out of a p-wave centrifugal
barrier. The lifetimes on the quasi-bound side of the resonance are
well reproduced by a standard multichannel scattering calculation,
(see Fig. \ref{lifetimes}), provided that partial waves with $l=3$
are included as well as those with $l=1$. The lifetimes so obtained
are about $1.6$ times what would be predicted by a single-channel
calculation using the triplet potential energy surface. This is
because, in the multichannel case, the atoms spend part of their
time in higher-lying channels and thus have fewer opportunities to
tunnel.

On the bound molecule side, we observe that for magnetic-field
detunings as large as 1 G from the resonance, the lifetimes are
independent of magnetic field and therefore binding energy. The
bound $m_l=0$ molecule lifetime is measured to be $2.3\pm0.2$  ms,
while the lifetime of molecules created on the $m_l=\pm1$ resonance
is measured to be $1.0\pm0.1$ ms. Our multichannel scattering model
predicts magnetic-field independent lifetimes set by dipolar
relaxation rates. The predicted lifetimes are $8.7$ ms for $m_l = 0$
state, $6.8$ ms for $m_l = 1$ state, and $1.5$ ms for $m_l = -1$
state. The shorter measured lifetimes may be explained by molecular
collisions, however, we have not yet been able to obtain
experimental evidence for a density dependence. The reason for this
discrepancy thus remains a challenge for future investigation.

This work has focused on the study of p-wave Feshbach pairs away
from the resonance where they are weakly coupled to free atoms. In
this regime, we have measured the pair lifetimes and binding
energies and, with the exception of the bound molecule lifetimes,
find good agreement with two-body theory.  We believe that the
techniques and results presented here constitute a starting point
for attempts to study the many-body properties of these gases near
the resonance.  Indeed, our data showing pair creation on the
resonance represents a first step in studying the behavior of a
quantum degenerate Fermi gas in the presence of resonant p-wave
interactions.

Unfortunately, in a $^{40}$K experiment, the short lifetimes
measured for the p-wave molecules make it unlikely that one could
make p-wave molecular condensates. However, it may be possible to
explore many-body physics on the quasi-bound side of the resonance
where p-wave interactions are strongly enhanced and inelastic decay
rates can be much slower than the tunneling rate. Furthermore, our
studies suggest that longer lifetimes may be possible in other
systems if a p-wave resonance occurs at low magnetic field where
dipolar relaxation rates could be suppressed or in the lowest Zeeman
state of the atoms where dipolar relaxation would be energetically
forbidden.

\begin{acknowledgments}
We acknowledge funding from the NSF and NASA.  We thank both the
JILA BEC group and Leo Radzihovsky for stimulating discussions.
\end{acknowledgments}



\begin{thebibliography}{10}
\bibitem{widths} In order to convert our width measurements to lifetimes we
have subtracted out the contribution to the measured width due to
the finite energy distribution of the atoms. The size of this
correction is approximately $15 \%$.
\end{thebibliography}

\end{document}